\documentclass[12pt]{article}
\usepackage[english]{babel}
\usepackage{amsmath,amsthm,amssymb,amsfonts,latexsym,url}
\usepackage{color}
\newtheorem{theorem}{Theorem}

\newtheorem{definition}{Definition}
\newtheorem{remark}{Remark}
\newtheorem{example}{Example}

\usepackage{mathptmx}      
\newcommand{\sech}{\,\mathrm{sech}}
\begin{document}
\title{A Consistent Approach to Approximate Lie Symmetries of Differential Equations}
\author{R.~D.~Salvo, M.~Gorgone and F.~Oliveri\\
\ \\
{\footnotesize Department of Mathematical and Computer Sciences,}\\
{\footnotesize Physical Sciences and Earth Sciences, University of Messina}\\
{\footnotesize Viale F. Stagno d'Alcontres 31, 98166 Messina, Italy}\\
{\footnotesize rdisalvo@unime.it; mgorgone@unime.it; foliveri@unime.it}
}

\date{Published in \textit{Nonlinear Dyn.} \textbf{91}, 371--386 (2018).}

\maketitle

\begin{abstract}
Lie theory of continuous transformations  provides a unified and powerful approach for handling differential 
equations.  Unfortunately, any small perturbation of an equation usually destroys some important symmetries, and this 
reduces the applicability of Lie group methods to differential equations arising in concrete applications. On the other 
hand, differential equations containing \emph{small terms} are commonly and successfully investigated by means of 
perturbative techniques. 
Therefore, it is desirable to combine Lie group methods with perturbation analysis, \emph{i.e.}, to  establish an 
approximate symmetry theory. There are two widely used approaches to approximate symmetries: the one proposed in 
1988 by Baikov, Gazizov and Ibragimov, and the one introduced
in 1989 by Fushchich and Shtelen. Moreover, some variations of the Fushchich--Shtelen method have been proposed 
with the aim of reducing the length of computations. Here, we propose a new approach that is consistent with 
perturbation theory and allows to extend all the relevant features of Lie group analysis to an approximate context. 
Some applications are also presented.
\end{abstract}

\noindent
\textbf{Keywords.} Differential equations; Approximate Lie symmetries; Perturbation techniques

\section{Introduction}
The study of Lie continuous transformation groups of differential equations  
\cite{Ovsiannikov,Ibragimov,Olver,Ibragimov:CRC,Olver2,Baumann,BlumanAnco,Meleshko2005,BlumanCheviakovAnco,Bordag}
represents a general algorithmic approach to  differential equations that, besides its intrinsic theoretical interest 
(\emph{e.g.}, for classifying differential equations according to the admitted symmetries), 
plays a fundamental role from a geometrical or analytical viewpoint. 
In fact, the exploitation of symmetries of differential equations often leads to their simplification.

For ordinary differential equations, the knowledge of their symmetries allows us to algorithmically 
lower their order or, possibly, reduce them to quadrature.
For partial differential equations, the symmetries suggest how to combine the dependent
and independent variables in such a way a reduction of their  
dimension arises, and special (invariant) solutions  
\cite{Olver,Ibragimov:CRC,OliveriSpeciale1998,OliveriSpeciale1999,OliveriSpeciale2002,Oliveri_ND2005,OliveriSpeciale2005} 
of initial and boundary value problems can be determined; symmetries play also a role in deriving conserved quantities, 
or in the algorithmic construction of invertible point transformations linking different differential
equations that turn out to be equivalent
\cite{BlumanCheviakovAnco,BlumanKumei:book,KumeiBluman,DonatoOliveri1993,DonatoOliveri1994,DonatoOliveri1995,DonatoOliveri1996,CurroOliveri2008,Oliveri2010,Oliveri2012,GorgoneOliveri_RM,GorgoneOliveri_JGP2017}.

Unfortunately,  any small perturbation in  a differential equation has the --- often dramatic --- effect of destroying many 
useful symmetries, and this limits the applicability of Lie group methods to concrete problems where equations 
involving terms of different order of magnitude may occur.  To overcome this inconvenient, some \emph{approximate 
symmetry theories} have been proposed in order to deal with differential equations involving small terms, and the 
notion of  \emph{approximate invariance} has been introduced.

The first, illuminating, paper was by Baikov, Gazizov and Ibragimov \cite{BGI-1989} (see also \cite{IbragimovKovalev}), 
who in 1988 proposed to expand in a perturbation series the Lie generator in order to have
an approximate generator.  The resulting theory is quite elegant since all the useful properties of \emph{exact} Lie 
symmetries can be adapted in the approximate sense: reduction of order of ordinary differential equations, 
approximately invariant solutions, approximate conservation laws, etc. Since its introduction, this approach has been 
applied to many physical models 
\cite{Wiltshire1996,Kovalev_ND2000,BaikovKordyukova2003,DolapciPakdemirli2004,Pakdemirli2004,%
IbragimovUnalJogreus2004,Wiltshire2006,Kara_ND2008,GazizovIbragimovLukashchuk2010,Gan_Qu_ND2010,%
GazizovIbragimov2014}. 
Nevertheless, the expanded generator is not consistent with the principles of perturbation analysis 
\cite{Nayfeh} because the dependent variables are not expanded. This implies that in several examples the 
approximately invariant solutions that are found with this method are not the most general ones. In 
\cite{DonatoPalumbo1,DonatoPalumbo2},  the approach proposed in \cite{BGI-1989} has been applied to the 
investigation of asymptotic waves in dissipative systems 
by expanding in a perturbation series both the Lie generator and the dependent variables.

In 1989, Fushchich and Shtelen \cite{FS-1989} proposed a different approach. The dependent variables are expanded 
in a  series as done in usual perturbation analysis;
terms are then separated at each order of approximation, and a system of equations to be solved in a hierarchy is 
obtained.  This resulting system is assumed to be coupled, and the approximate symmetries of the original equations 
are defined as the \emph{exact symmetries} of the equations obtained from separation.
This approach has an obvious simple and coherent basis. \emph{Per contra}, a lot of algebra (especially for higher
order perturbations) is needed; moreover, the basic assumption of a fully coupled system is too strong, since
the equations at a level are not influenced by those at higher levels. In addition, there is no possibility to work in a 
hierarchy: for instance, if one computes first order approximate symmetries, and then searches for second order 
approximate symmetries, all the work must be done from the very beginning. Applications of this method to various 
equations can be found, for instance, in the papers 
\cite{Wiltshire2006,Euler1,Euler2,Euler3,Diatta}.

In 2004, Pakdemirli \emph{et al.} \cite{Pakdemirli2004} compared the two different approaches, and proposed a third 
method as a variation of Fushchich--Shtelen one by removing 
the assumption of a fully coupled system. This method applies to differential equations having the form
\begin{equation}
\label{pakdemirli}
\mathcal{L}(\mathbf{u})+\varepsilon\mathcal{N}(\mathbf{u})=0,
\end{equation}
where $\mathcal{L}(\mathbf{u})$ and $\mathcal{N}(\mathbf{u})$ are a linear and a nonlinear differential operator, 
respectively. Expanding the dependent variables,
\begin{equation}
\mathbf{u}=\mathbf{u}_{(0)}+\varepsilon \mathbf{u}_{(1)}+\varepsilon^2\mathbf{u}_{(2)}+\ldots
\end{equation}
equation~\eqref{pakdemirli} provides
\begin{equation}
\label{hier}
\begin{aligned}
&\mathcal{L}(\mathbf{u}_{(0)})=0=h_0(\mathbf{x}),\\
&\mathcal{L}(\mathbf{u}_{(1)})=-\mathcal{N}(\mathbf{u}_{(0)})=h_1(\mathbf{x}),\\
&\mathcal{L}(\mathbf{u}_{(2)})=-\mathcal{N}(\mathbf{u}_{(0)},\mathbf{u}_{(1)})=h_2(\mathbf{x}),\\
&\ldots
\end{aligned}
\end{equation}
Since the differential operators in the left--hand sides of all the equations are the same, and the right--hand sides can 
be considered as arbitrary functions of the independent variables which are determined sequentially starting from the 
first equation, the approximate symmetries of the nonlinear equation~\eqref{pakdemirli}
are defined as the exact symmetries of  the linear nonhomogeneous equation
\begin{equation}
\label{symbolhier}
\mathcal{L}(\mathbf{u}) = h(\mathbf{x}),
\end{equation}
with $h(\mathbf{x})$ considered as an arbitrary function.
Due to the circumstance that the functions $h_i(\mathbf{x})$ in \eqref{hier} are in fact known functions, substituting 
these forms into the general expression of the
symmetries of equation~\eqref{symbolhier}, one obtains the symmetries at each level of approximation.
The involved  algebra is much less than that required by Fushchich--Shtelen method, and it is
possible to work in a hierarchy; nevertheless, the method is not general.

It is well known that searching for Lie symmetries of differential equations requires a lot of long and tedious, though 
straightforward, calculations. Nevertheless, many symbolic packages do exist doing almost
automatically the required work (see \cite{Ibragimov:CRC,Baumann,Hereman,Butcher-Carminati-Vu,Cheviakov2010}, 
and references therein).  The computation of approximate Lie symmetries of differential equations imposes additional 
computational costs. In a recent paper \cite{Jefferson-Carminati}, Jefferson and Carminati  described the MAPLE 
package ASP (Automated Symmetry Package), which is an add-on to the MAPLE symmetry package DESOLVII 
\cite{Vu-Jefferson-Carminati}, for automating the above three methods of determining approximate symmetries for 
differential equations.

A further variant of Fushchich--Shtelen method has been proposed in \cite{Valenti} for the analysis of a third order
partial differential equation describing one--dimensional wave propagation in nonlinear
dissipative media:
\begin{equation}
\label{valenti}
w_{tt}-f(w_x)w_{xx}=\varepsilon w_{xxt},
\end{equation}
where $f(w_x)$ is an arbitrary function and the subscripts $t$ and $x$ denote partial derivatives. Inserting the expansion 
of the dependent variable in power series of $\varepsilon$,
\[
w(t,x,\varepsilon)=w_0(t,x)+\varepsilon w_1(t,x)+O(\varepsilon^2),
\]
into equation~\eqref{valenti}, and separating terms with different powers in $\varepsilon$, an approximate system is 
found. Then, the approximate symmetries of equation \eqref{valenti} are defined as the (exact) symmetries of the 
approximate system through the Lie group generated by
\[
\begin{aligned}
\Xi&=\xi^1_0(t,x,w_0)\frac{\partial}{\partial t} + \xi^2_0(t,x,w_0)\frac{\partial}{\partial x}+\eta_0(t,x,w_0)
\frac{\partial}{\partial w_0}\\
&+\left(\eta_{10}(t,x,w_0)+\eta_{11}(t,x,w_0)w_1\right)\frac{\partial}{\partial w_1}.
\end{aligned}
\]
This approach allows to work in a hierarchy; \emph{per contra},
there is not a full mix between Lie group methods and perturbation analysis (as in the case of Fushchich--Shtelen method). 
Applications of this approach can be found in \cite{Ruggieri-Speciale2016,Ruggieri-Speciale2017}.

The aim of this paper is to propose an \emph{approximate symmetry theory} 
which is consistent with perturbative analysis and inherits the relevant properties of exact Lie symmetries of differential equations. More precisely, 
the dependent variables are expanded in power series of the small parameter as done in classical perturbative 
analysis; then, instead of considering the approximate symmetries as the exact symmetries of the approximate system 
(as done  in Fushchich--Shtelen method), the consequent expansion of the Lie generator is constructed, and the 
approximate invariance  with respect to the approximate Lie generator is introduced, as in 
Baikov--Gazizov--Ibragimov method. 
Of course, the method requires more computations than that required for determining exact Lie 
symmetries; nevertheless, a general Reduce \cite{Reduce} package (ReLie, \cite{Relie}), written by one of the 
authors (F.~O.), is able to compute -- besides exact, conditional and contact symmetries as well as equivalence 
transformations of differential equations -- approximate Lie symmetries as introduced in this paper.

The paper is organized as follows. Section~\ref{approximategroup} is devoted to fix the notation, define the 
approximate one--parameter Lie groups of transformations, and properly state all the relevant properties of classical 
Lie groups of transformations in the \emph{approximate framework}. Section~\ref{approximatesymmetries} deals 
more specifically with approximate Lie symmetries of differential equations. In Section~\ref{applications}, by means of 
some examples, the use of approximate Lie symmetries to lower the order of an ordinary differential 
equation, and determine approximately invariant solutions of partial differential equations is shown. Finally, 
Section~\ref{conclusions} contains our conclusions.

\section{Approximate one--parameter Lie groups}
\label{approximategroup}

We begin this Section by introducing the necessary notation in order to define approximate one--parameter Lie groups 
of transformations.

Let $f(\mathbf{z},\varepsilon)$ be a $C^\infty$ function depending on $\mathbf{z}\equiv(z_1,\ldots,z_N)\in D\subseteq 
\mathbb{R}^N$ and the small parameter $\varepsilon\in\mathbb{R}$; in the following, we will consider such kind of 
functions locally in a neighborhood of $\varepsilon=0$.

Let us expand $\mathbf{z}$ in power series of $\varepsilon$,
\begin{equation}
\mathbf{z}=\mathbf{z}_{(0)}+\varepsilon\mathbf{z}_{(1)}+\dots+\varepsilon^p\mathbf{z}_{(p)}+O(\varepsilon^{p+1}),
\end{equation}
with $\mathbf{z}_{(k)}\equiv(z_{(k)1},\ldots,z_{(k)N})$, whereupon
\begin{equation}
f(\mathbf{z},\varepsilon)=\sum_{k=0}^p\sum_{|\sigma|=k}
\frac{\varepsilon^{\sigma_0}}{\sigma_0!}\left(\prod_{i=1}^N \frac{(z_i-z_{(0)i})^{\sigma_i}}{\sigma_i!}\right)
\left.\frac{\partial^{|\sigma|}f(\mathbf{z},\varepsilon)}{\partial \varepsilon^{\sigma_0}\partial z_1^{\sigma_1}\cdots\partial 
z_N^{\sigma_N}}\right|_{\varepsilon=0}+O(\varepsilon^{p+1}),
\end{equation}
$\sigma$ being the multi--index $(\sigma_0,\sigma_1,\ldots,\sigma_N)$, and $|\sigma|=\sigma_0+\sigma_1+\dots+
\sigma_N$. 

By means of the positions
\begin{equation}
\begin{aligned}
&{f}_{(0)}(\mathbf{z}_{(0)})=\left.f(\mathbf{z},\varepsilon)\right|_{\varepsilon=0},\\
&{f}_{(k)}(\mathbf{z}_{(0)})=\left.\frac{\partial^k f(\mathbf{z},\varepsilon)}{\partial \varepsilon^k}\right|_{\varepsilon=0},
\end{aligned}
\end{equation}
since it is
\begin{equation}
\left.\frac{\partial^{|\sigma|}f(\mathbf{z},\varepsilon)}{\partial \varepsilon^{\sigma_0}\partial z_1^{\sigma_1}\cdots\partial 
z_N^{\sigma_N}}\right|_{\varepsilon=0}=\frac{\partial^{|\sigma|-\sigma_0} {f}_{(\sigma_0)}(\mathbf{z}_{(0)})}{\partial 
z_{(0)1}^{\sigma_1}\cdots\partial z_{(0)N}^{\sigma_N}},
\end{equation}
we have:
\begin{equation}
f(\mathbf{z},\varepsilon)=\sum_{k=0}^p\sum_{|\sigma|=k}
\frac{\varepsilon^{\sigma_0}}{\sigma_0!}\left(\prod_{i=1}^N \frac{(z_i-z_{(0)i})^{\sigma_i}}{\sigma_i!}\right)
\frac{\partial^{|\sigma|-\sigma_0}{f}_{(\sigma_0)}(\mathbf{z}_{(0)})}{\partial z_{(0)1}^{\sigma_1}\cdots\partial z_{(0)N}
^{\sigma_N}}+O(\varepsilon^{p+1});
\end{equation}
therefore, the expansion in power series of $\varepsilon$ is characterized (up to the order $p$ in $\varepsilon$) by 
$p+1$ functions of $\mathbf{z}_{(0)}$. Such an expansion can be written as
\begin{equation}
\begin{aligned}
&f(\mathbf{z},\varepsilon)=\sum_{k=0}^p \varepsilon^k \widetilde{f}_{(k)}(\mathbf{z}_{(0)},\ldots, \mathbf{z}_{(k)})
+O(\varepsilon^{p+1}),
\end{aligned}
\end{equation}
where $\widetilde{f}_{(k)}$ $(k>0)$ are suitable polynomials in $\mathbf{z}_{(1)},\ldots,\mathbf{z}_{(k)}$ with 
coefficients given by ${f}_{(0)}(\mathbf{z}_{(0)}),\ldots,{f}_{(k)}(\mathbf{z}_{(0)})$ and their derivatives with respect to 
$\mathbf{z}_{(0)}$. More precisely, the functions $\widetilde{f}_{(k)}$  are defined as follows:
\begin{equation}
\begin{aligned}
&\widetilde{f}_{(0)}={f}_{(0)},\\
&\widetilde{f}_{(k+1)}=\frac{1}{k+1}\mathcal{R}[\widetilde{f}_{(k)}],
\end{aligned}
\end{equation} 
$\mathcal{R}$ being a \emph{linear} recursion operator satisfying \emph{product rule} of derivatives and such that
\begin{equation}
\label{R_operator}
\begin{aligned}
&\mathcal{R}\left[\frac{\partial^{|\tau|}{f}_{(k)}(\mathbf{z}_{(0)})}{\partial z_{(0)1}^{\tau_1}\dots\partial z_{(0)N}^{\tau_N}}
\right]=\frac{\partial^{|\tau|}{f}_{(k+1)}(\mathbf{z}_{(0)})}{\partial z_{(0)1}^{\tau_1}\dots\partial z_{(0)N}^{\tau_N}}\\
&\phantom{\mathcal{R}\left[\frac{\partial^{|\tau|}{f}_{(k)}(\mathbf{z}_{(0)})}{\partial z_{(0)1}^{\tau_1}\dots\partial z_{(0)N}
^{\tau_N}}\right]}
+\sum_{i=1}^N\frac{\partial}{\partial z_{(0)i}}\left(\frac{\partial^{|\tau|} {f}_{(k)}(\mathbf{z}_{(0)})}{\partial z_{(0)1}
^{\tau_1}\dots\partial z_{(0)N}^{\tau_N}}\right)z_{(1)i},\\
&\mathcal{R}[z_{(k)j}]=(k+1)z_{(k+1)j},
\end{aligned}
\end{equation}
where $k\ge 0$,  $j=1,\ldots, N$, $|\tau|=\tau_1+\cdots+\tau_N$.

\begin{definition}
Given two smooth functions $f(\mathbf{z},\varepsilon)$ and $g(\mathbf{z},\varepsilon)$, and fixing a positive integer 
$p$, we will write
\begin{equation}
f(\mathbf{z},\varepsilon)\stackrel{p}{\approx} g(\mathbf{z},\varepsilon)
\end{equation} 
when it is
\begin{equation}
f(\mathbf{z},\varepsilon)= g(\mathbf{z},\varepsilon)+O(\varepsilon^{p+1}).
\end{equation}
This means that $g(\mathbf{z},\varepsilon)$ has the same Taylor expansion as $f(\mathbf{z},\varepsilon)$ up to the 
order $p$ in $\varepsilon$. In the rest of the paper, we shall use the simplest notation $\approx$ instead of 
$\stackrel{p}{\approx}$. 
\end{definition}

\begin{definition}[Approximate one--parameter Lie groups]
Let  $D\subseteq\mathbb{R}^N$ be an open domain, $E\subseteq\mathbb{R}$ an interval containing zero, 
$S\subseteq\mathbb{R}$ another interval (bounded or unbounded) containing zero and
such that $\forall\, a,b\in S,\;a+b\in S$. 
The set of transformations
\begin{equation}
\begin{aligned}
& \mathbf{Z}: D \times E\times S \rightarrow D,\\ 
&(\mathbf{z},\varepsilon,a)\mapsto \mathbf{z}^\star=\mathbf{Z}(\mathbf{z},\varepsilon ;a),
\end{aligned}
\end{equation}
depending on the parameter $a$, where $\varepsilon\in E$ is small, is a one--parameter $(a)$ 
Lie group of transformations if 
\begin{enumerate}
\item $\mathbf{Z}(\mathbf{Z}(\mathbf{z},\varepsilon;a),\varepsilon;b)=\mathbf{Z}(\mathbf{z},\varepsilon;a+b)$;
\item $\mathbf{Z}(\mathbf{z},\varepsilon;0)=\mathbf{z}$;
\item for each value of the parameter $a\in S$ and each $\varepsilon\in E$ the transformations are one--to--one onto 
$D$;
\item $\mathbf{Z}$ is $C^\infty$ with respect to $\mathbf{z}\in D$ and $\varepsilon\in E$, and analytic with respect to 
the parameter $a$.
\end{enumerate}
Let us fix a positive integer $p$, and expand $\mathbf{z}$ in power series of $\varepsilon$,
\begin{equation}
\mathbf{z}=\mathbf{z}_{(0)}+\varepsilon\mathbf{z}_{(1)}+\dots+\varepsilon^p\mathbf{z}_{(p)}+O(\varepsilon^{p+1}),
\end{equation}
whereupon, by defining
\begin{equation}
\begin{aligned}
&{\mathbf{Z}}_{(0)}(\mathbf{z}_{(0)})=\left.\mathbf{Z}(\mathbf{z},\varepsilon)\right|_{\varepsilon=0},\\
&{\mathbf{Z}}_{(k)}(\mathbf{z}_{(0)})=\left.\frac{\partial^k \mathbf{Z}(\mathbf{z},\varepsilon)}{\partial \varepsilon^k}\right|
_{\varepsilon=0},
\end{aligned}
\end{equation}
we have
\begin{equation}
\mathbf{Z}(\mathbf{z},\varepsilon;a)=\sum_{k=0}^p \varepsilon^k \widetilde{\mathbf{Z}}_{(k)}(\mathbf{z}_{(0)},\ldots, 
\mathbf{z}_{(k)};a)+O(\varepsilon^{p+1}),
\end{equation}
along with
\begin{equation}
\begin{aligned}
&\widetilde{\mathbf{Z}}_{(0)}(\mathbf{z}_{(0)})={\mathbf{Z}}_{(0)}(\mathbf{z}_{(0)}),\\
&\widetilde{\mathbf{Z}}_{(k+1)}=\frac{1}{k+1}\mathcal{R}[\widetilde{\mathbf{Z}}_{(k)}], \qquad k\ge 0,
\end{aligned}
\end{equation} 
where $\mathcal{R}$ is the recursion operator defined in \eqref{R_operator}.

The transformation
\begin{equation}
\mathbf{z}^\star=\mathbf{Z}(\mathbf{z},\varepsilon;a)\approx\sum_{k=0}^p \varepsilon^k\widetilde{\mathbf{Z}}_{(k)}
(\mathbf{z}_{(0)},\ldots, \mathbf{z}_{(k)};a)
\end{equation} 
is an approximate one--parameter Lie group with respect to the parameter $a$  if
\begin{equation}
\mathbf{Z}(\mathbf{z},\varepsilon;0)\approx\mathbf{z},\qquad
\mathbf{Z}(\mathbf{Z}(\mathbf{z},\varepsilon;a),\varepsilon;b)\approx\mathbf{Z}(\mathbf{z},\varepsilon;a+b),
\end{equation}
and the condition $\mathbf{Z}(\mathbf{z},\varepsilon;a)\approx\mathbf{z}$ for all $\mathbf{z}$ implies $a=0$.
\end{definition}

Hereafter we are going to show that all the relevant features of classical Lie groups of transformations can be extended in the 
approximate sense.

\begin{definition}[Generator of an approximate Lie group]
The generator of the  Lie group of transformations
\begin{equation}\label{transf}
\mathbf{z}^\star=\mathbf{Z}(\mathbf{z},\varepsilon;a)\approx\sum_{k=0}^p \varepsilon^k\widetilde{\mathbf{Z}}_{(k)}
(\mathbf{z}_{(0)},\ldots, \mathbf{z}_{(k)};a)
\end{equation} 
is given by
\begin{equation}
\Xi=\sum_{i=1}^N \zeta_i(\mathbf{z},\varepsilon)\frac{\partial}{\partial z_i},
\end{equation}
where
\begin{equation}
\zeta_i(\mathbf{z},\varepsilon)= \left.\frac{\partial Z_i(\mathbf{z},\varepsilon;a)}{\partial a}\right|_{a=0}.
\end{equation}
Inserting the expansion of $\mathbf{z}$, we have 
\begin{equation}
\zeta_i(\mathbf{z},\varepsilon)\approx \sum_{k=0}^p \varepsilon^k\widetilde{\zeta}_{(k)i}(\mathbf{z}_{(0)},\ldots,\mathbf{z}_{(k)})
\end{equation}
where
\begin{equation}
\widetilde{\zeta}_{(k)i}(\mathbf{z}_{(0)},\ldots,\mathbf{z}_{(k)})=\left.\frac{\partial \widetilde{Z}_{(k)i}(\mathbf{z}_{(0)},
\ldots, \mathbf{z}_{(k)};a)}{\partial a}\right|_{a=0}.
\end{equation}
Setting
\begin{equation}
\begin{aligned}
&{\zeta}_{(0)i}(\mathbf{z}_{(0)})= \left.\zeta_i(\mathbf{z},\varepsilon)\right|_{\varepsilon=0},\\
&{\zeta}_{(k)i}(\mathbf{z}_{(0)})= \left.\frac{\partial^k \zeta_i(\mathbf{z},\varepsilon)}{\partial\varepsilon^k}\right|
_{\varepsilon=0},
\end{aligned}
\end{equation}
it is
\begin{equation}
\begin{aligned}
&\widetilde{\zeta}_{(0)i}(\mathbf{z}_{(0)})={\zeta}_{(0)i}(\mathbf{z}_{(0)}),\\
&\widetilde{\zeta}_{(k+1)i}= \frac{1}{k+1}\mathcal{R}[\widetilde{\zeta}_{(k)i}],
\end{aligned}
\end{equation}
where $\mathcal{R}$ is the recursion operator defined in \eqref{R_operator}.
Therefore, we may write the approximate Lie generator as
\begin{equation}
\Xi\approx \sum_{k=0}^p\varepsilon^k\widetilde{\Xi}_{(k)},
\end{equation}
with obvious meaning of terms.
\end{definition}

\begin{example}
If $p=1$, the generator of the approximate Lie group of transformations reads
\begin{equation}
\Xi\approx  \sum_{i=1}^N \left({\zeta}_{(0)i}(\mathbf{z}_0)+\varepsilon\left({\zeta}_{(1)i}(\mathbf{z}_0)+\sum_{j=1}
^N\frac{\partial {\zeta}_{(0)i}(\mathbf{z}_0)}{\partial z_{(0)j}}z_{(1)j}\right)\right)\frac{\partial}{\partial z_i}.
\end{equation}
\end{example}

\begin{definition}[Approximate Lie's equations]
Let
\begin{equation}
\Xi\approx \sum_{k=0}^p\varepsilon^k\widetilde{\Xi}_{(k)}
\end{equation}
be an approximate Lie generator; the corresponding finite transformation 
\begin{equation}
\mathbf{z}^\star\approx\sum_{k=0}^p \varepsilon^k\widetilde{\mathbf{Z}}_{(k)}(\mathbf{z}_{(0)},\ldots, 
\mathbf{z}_{(k)};a)
\end{equation}
is recovered by solving the approximate Lie's equations
\begin{equation}
\begin{aligned}
&\frac{d\mathbf{z}_{(0)}^\star}{da}=\widetilde{\boldsymbol \zeta}_{(0)}(\mathbf{z}_{(0)}^\star),\quad &&\mathbf{z}_{(0)}
^\star(0)= \mathbf{z}_{(0)},\\
&\frac{d\mathbf{z}_{(k)}^\star}{da}=\widetilde{\boldsymbol \zeta}_{(k)}(\mathbf{z}_{(0)}^\star,\ldots,\mathbf{z}_{(k)}
^\star),
\quad &&\mathbf{z}_{(k)}^\star(0)= \mathbf{z}_{(k)}, \qquad
k=1,\ldots,p.
\end{aligned}
\end{equation}
\end{definition}

\begin{definition}[Invariance of a function]
A smooth function
\begin{equation}
F(\mathbf{z},\varepsilon)\approx \sum_{k=0}^p\varepsilon^k\widetilde{F}_{(k)}(\mathbf{z}_{(0)},\ldots,\mathbf{z}_{(k)})
\end{equation}
is approximately invariant with respect to 
\begin{equation}
\label{transf1}
\mathbf{z}^\star\approx\sum_{k=0}^p \varepsilon^k\widetilde{\mathbf{Z}}_{(k)}(\mathbf{z}_{(0)},\ldots, \mathbf{z}
_{(k)};a)
\end{equation}
if
\begin{equation}
F(\mathbf{z}^\star,\varepsilon)\approx F(\mathbf{z},\varepsilon) .
\end{equation}
\end{definition}

By using the same arguments as those in classical Lie theory, the following theorem about the approximate invariance 
of a function can be stated.
\begin{theorem}\label{thm:invariance}
Let
\begin{equation}
\Xi\approx\sum_{k=0}^p\varepsilon^k\widetilde{\Xi}_{(k)}
\end{equation}
be an approximate Lie generator of \eqref{transf1}.
Then, the function
\begin{equation}
F(\mathbf{z},\varepsilon)\approx \sum_{k=0}^p\varepsilon^k\widetilde{F}_{(k)}(\mathbf{z}_{(0)},\ldots,\mathbf{z}_{(k)})
\end{equation}
is approximately invariant with respect to \eqref{transf1} if and only if
\begin{equation}
\label{cond_inv}
\Xi(F(\mathbf{z},\varepsilon))\approx 0 .
\end{equation}
\end{theorem}

Some simple algebra to derive the constraints arising from condition \eqref{cond_inv} is worth of
being explicitly given. It is:
\begin{equation}
\label{effectiveoperator}
\begin{aligned}
&\Xi(F(\mathbf{z},\varepsilon))\approx\sum_{j=0}^p\varepsilon^j\left(\sum_{i=1}^N \widetilde{\zeta}_{(j)i}(\mathbf{z}
_{(0)},\ldots,\mathbf{z}_{(j)})\frac{\partial F(\mathbf{z},\varepsilon)}{\partial z_i}\right)\approx\\
&\qquad\approx\sum_{j,k=0}^p\varepsilon^{j+k}\left(\sum_{i=1}^N \widetilde{\zeta}_{(j)i}(\mathbf{z}_{(0)},\ldots,\mathbf{z}_{(j)})\frac{\partial \widetilde{F}_{(k)}(\mathbf{z}_{(0)},\ldots,\mathbf{z}_{(k)})}{\partial z_{(0)i}}\right)\approx\\
&\qquad\approx\sum_{k=0}^p\varepsilon^{k}\left(\sum_{j=0}^k\sum_{i=1}^N\widetilde{\zeta}_{(j)i}(\mathbf{z}_{(0)},\ldots,\mathbf{z}_{(j)}) \frac{\partial \widetilde{F}_{(k-j)}(\mathbf{z}_{(0)},\ldots,\mathbf{z}_{(k-j)})}{\partial z_{(0)i}}\right).
\end{aligned}
\end{equation}

From \eqref{cond_inv}, taking into account \eqref{effectiveoperator}, and separating the various 
powers in $\varepsilon$, we easily obtain the conditions
\begin{equation}
\label{reduced_cond}
\sum_{j=0}^k\Xi_{(j)}{F}_{(k-j)}(\mathbf{z}_{(0)}))=\sum_{j=0}^k\sum_{i=1}^N{\zeta}_{(j)i}(\mathbf{z}_{(0)}) 
\frac{\partial {F}_{(k-j)}(\mathbf{z}_{(0)})}{\partial z_{(0)i}}=0, 
\end{equation}
for $k=0,\ldots,p$, where we introduced  the \emph{reduced operators}
\begin{equation}
\Xi_{(j)}=\sum_{i=1}^N\zeta_{(j)i}(\mathbf{z}_0)\frac{\partial}{\partial z_{(0)i}}, \qquad j=0,\ldots, p.
\end{equation}
Therefore, the approximate invariance condition of a function $F(\mathbf{z},\varepsilon)$ yields 
the $p+1$ conditions~\eqref{reduced_cond} for the $p+1$ functions characterizing the expansion (up to the order $p$ 
in $\varepsilon$) of the function $F$.

\begin{example}
Let $p=1$, and $\mathbf{z}\equiv(z_1,z_2)$; a smooth function 
\[
\begin{aligned}
F(\mathbf{z},\varepsilon)&\approx 
F_0(z_{(0)1},z_{(0)2})\\
&+\varepsilon\left(F_1(z_{(0)1},z_{(0)2})+\frac{\partial F_0(z_{(0)1},z_{(0)2})}{\partial z_{(0)1} }z_{(1)1}+\frac{\partial 
F_0(z_{(0)1},z_{(0)2})}{\partial z_{(0)2} }z_{(1)2}\right),  
\end{aligned}
\]
is approximately invariant with respect to the Lie generator
\begin{equation}
\Xi\approx \left(z_{(0)2}+\varepsilon z_{(1)2}\right)\frac{\partial}{\partial z_1}-\left(z_{(0)1}+\varepsilon z_{(1)1}\right)
\frac{\partial}{\partial z_2},
\end{equation} 
provided that
\begin{equation}
\begin{aligned}
&z_{(0)2}\frac{\partial F_0(z_{(0)1},z_{(0)2})}{\partial z_{(0)1}}-z_{(0)1}\frac{\partial F_0(z_{(0)1},z_{(0)2})}{\partial 
z_{(0)2}}=0,\\
&z_{(0)2}\frac{\partial F_1(z_{(0)1},z_{(0)2})}{\partial z_{(0)1}}-z_{(0)1}\frac{\partial F_1(z_{(0)1},z_{(0)2})}{\partial 
z_{(0)2}}=0,
\end{aligned}
\end{equation}
whereupon it is
\begin{equation}
F(z_1,z_2)\approx  F_0(r_0^2)+\varepsilon\left(F_1(r_0^2)+
2(z_{(0)1}z_{(1)1}+z_{(0)2}z_{(1)2})F_0^\prime(r_0^2)\right),
\end{equation}
$F_0$ and $F_1$ being arbitrary functions of $r_0^2=z_{(0)1}^2+z_{(0)2}^2$, and the prime ${}^\prime$ 
denoting the differentiation with respect to the argument.
\end{example}

\begin{definition}[Invariance of an equation]
The equation
\begin{equation}
F(\mathbf{z},\varepsilon)\approx \sum_{k=0}^p\varepsilon^k\widetilde{F}_{(k)}(\mathbf{z}_{(0)},\ldots,\mathbf{z}
_{(k)})=0
\end{equation}
is approximately invariant with respect to \eqref{transf1} if
\begin{equation}
F(\mathbf{z}^\star,\varepsilon)\approx 0\quad \hbox{when}\quad F(\mathbf{z},\varepsilon)\approx 0 .
\end{equation}
\end{definition}

\begin{theorem}
Let
\begin{equation}
\Xi\approx\sum_{k=0}^p\varepsilon^k\widetilde{\Xi}_{(k)}
\end{equation}
be an approximate Lie generator of \eqref{transf1}.

Then, the equation
\begin{equation}
F(\mathbf{z},\varepsilon)\approx \sum_{k=0}^p\varepsilon^k\widetilde{F}_{(k)}(\mathbf{z}_{(0)},\ldots,\mathbf{z}
_{(k)})=0
\end{equation}
is approximately invariant with respect to \eqref{transf1} if and only if
\begin{equation}
\Xi(F(\mathbf{z},\varepsilon))\approx 0 \quad \hbox{when}\quad F(\mathbf{z},\varepsilon)\approx 0,
\end{equation}
\emph{i.e.},
\begin{equation}
\Xi(F(\mathbf{z},\varepsilon))\approx\lambda(\mathbf{z},\varepsilon) F(\mathbf{z},\varepsilon),
\end{equation}
$\lambda(\mathbf{z,\varepsilon})$ being a multiplier.
\end{theorem}

\begin{definition}[Approximate canonical variables]
Let
\begin{equation}
\Xi\approx\sum_{k=0}^p\varepsilon^k\widetilde\Xi_{(k)}
\end{equation}
be an approximate Lie generator of \eqref{transf1}.

Then,
\begin{equation}
\mathbf{w}(\mathbf{z},\varepsilon)\approx \sum_{k=0}^{p}\varepsilon^k\widetilde{\mathbf{w}}_{(k)}(\mathbf{z}_{(0)},
\ldots,\mathbf{z}_{(k)})
\end{equation}
gives a set of approximate canonical variables if
\begin{equation}
\begin{aligned}
&\Xi(w_i)\approx 0, \qquad i=1,\ldots,N-1,\\
&\Xi(w_N)\approx 1.
\end{aligned}
\end{equation}
In terms of the approximate canonical variables, the approximate Lie generator writes as
\begin{equation}
\Xi\approx \frac{\partial}{\partial w_N},
\end{equation}
and the approximate Lie group of transformations corresponds to the translation of the variable $w_N$ only.
\end{definition}

\begin{example}
Let $\mathbf{z}\equiv(z_1,z_2)\in \mathbb{R}^2$, $p=1$ and 
\begin{equation}
\Xi\approx 
\left(
z_{(0)1}^2+\varepsilon(2z_{(0)1}z_{(1)1})\right)\frac{\partial}{\partial z_1}
+\left(z_{(0)1}z_{(0)2}+\varepsilon(z_{(0)1}z_{(1)2}+z_{(0)2}z_{(1)1})\right)\frac{\partial}{\partial z_2}.
\end{equation}
The first order approximate canonical variables
\begin{equation}
\mathbf{w}(\mathbf{z},\varepsilon)\approx {\mathbf{w}}_{(0)}(\mathbf{z}_{(0)})+\varepsilon\left({\mathbf{w}}_{(1)}
(\mathbf{z}_{(0)})+\sum_{i=1}^2\frac{\partial {\mathbf{w}}_{(0)}(\mathbf{z}_{(0)})}{\partial z_{(0)i}}z_{(1)i}\right)
\end{equation}
are determined by solving the following system of partial differential equations:
\[
\begin{aligned}
&z_{(0)1}^2\frac{\partial \widetilde{w}_{(0)1}}{\partial z_{(0)1}}+z_{(0)1}z_{(0)2}\frac{\partial \widetilde{w}_{(0)1}}{\partial 
z_{(0)2}}=0,\quad
&&z_{(0)1}^2\frac{\partial \widetilde{w}_{(0)2}}{\partial z_{(0)1}}+z_{(0)1}z_{(0)2}\frac{\partial \widetilde{w}_{(0)2}}
{\partial z_{(0)2}}=1,\\
&z_{(0)1}^2\frac{\partial \widetilde{w}_{(1)1}}{\partial z_{(0)1}}+z_{(0)1}z_{(0)2}\frac{\partial \widetilde{w}_{(1)1}}{\partial 
z_{(0)2}}=0,\quad
&&z_{(0)1}^2\frac{\partial \widetilde{w}_{(1)2}}{\partial z_{(0)1}}+z_{(0)1}z_{(0)2}\frac{\partial \widetilde{w}_{(1)2}}
{\partial z_{(0)2}}=0,
\end{aligned}
\]
whereupon we have
\[
\begin{aligned}
w_1&\approx\frac{z_{(0)2}}{z_{(0)1}}+\varepsilon\left(\frac{z_{(0)2}}{z_{(0)1}}-\frac{z_{(0)2} z_{(1)1}}{z_{(0)1}^2}+
\frac{z_{(1)2}}{z_{(0)1}}\right),\\
w_2&\approx-\frac{1}{z_{(0)1}}+\varepsilon\left(\frac{z_{(0)2}}{z_{(0)1}}+\frac{z_{(1)1}}{z_{(0)1}^2}\right),
\end{aligned}
\]
and the Lie generator reduces to
\[
\Xi\approx \frac{\partial}{\partial w_2}.
\] 
\end{example}

\begin{definition}[Approximate Lie bracket]
Let
\begin{equation}
\begin{aligned}
&\Xi_1\approx\sum_{k=0}^p\varepsilon^k \widetilde{\Xi}_{1(k)}, \qquad &&\Xi_2\approx\sum_{k=0}^p\varepsilon^k 
\widetilde{\Xi}_{2(k)},\\
&\widetilde{\Xi}_{1(k)}=\sum_{j=1}^N \widetilde{\zeta}_{1(k)j}\frac{\partial}{\partial z_j},\qquad
&&\widetilde{\Xi}_{2(k)}=\sum_{j=1}^N \widetilde{\zeta}_{2(k)j}\frac{\partial}{\partial z_j}
\end{aligned}
\end{equation}
be two approximate Lie generators. Their approximate Lie bracket is defined as follows:
\[
\begin{aligned}
&[\Xi_1,\Xi_2] \approx \sum_{k=0}^p \varepsilon^k\left(\sum_{i=0}^k\left[\widetilde{\Xi}_{1(i)},\widetilde{\Xi}_{2(k-i)}\right]
\right)\approx\\
&\approx\sum_{k=0}^p \varepsilon^k\left(\sum_{i=0}^k\left(\widetilde{\Xi}_{1(i)}\,\widetilde{\Xi}_{2(k-i)}-\widetilde{\Xi}
_{2(k-i)}\,\widetilde{\Xi}_{1(i)}\right)\right)\approx\\
&\approx\sum_{k=0}^p\varepsilon^k\left(\sum_{i=0}^k\left(\sum_{m=1}^N\left(\sum_{j=1}^N\left(\widetilde{\zeta}_{1(i)j}
\frac{\partial \widetilde{\zeta}_{2(k-i)m}}{\partial z_{(0)j}}-\widetilde{\zeta}_{2(k-i)j}\frac{\partial \widetilde{\zeta}_{1(i)m}}
{\partial z_{(0)j}}\right)\frac{\partial}{\partial z_{m}}\right)\right)\right).
\end{aligned}
\]
\end{definition}

The definition of approximate Lie bracket allows us to introduce a structure of approximate Lie algebra, and the 
following two theorems can be easily proved.

\begin{theorem}
The set of approximate Lie generators leaving a function approximately invariant has the structure of an approximate 
Lie algebra.
\end{theorem}

\begin{theorem}
The set of approximate Lie generators leaving an equation approximately invariant has the structure of an approximate 
Lie algebra.
\end{theorem}

\section{Approximate symmetries of differential equations}
\label{approximatesymmetries}
Here we use the results of previous Section in order to define and compute the approximate Lie symmetries of 
differential equations involving small terms. Within this framework, we distinguish the independent variables $
\mathbf{x}$ from the dependent ones $\mathbf{u}$, and limit ourselves to expand only the dependent variables in 
powers of $\varepsilon$.

Let 
\begin{equation}
\Delta(\mathbf{x},\mathbf{u},\mathbf{u}^{(r)};\varepsilon)=0
\end{equation}
be a differential equation of order $r$, where $\mathbf{u}^{(r)}$ denotes the set of all derivatives of the dependent 
variables $\mathbf{u}\in U\subseteq\mathbb{R}^m$ with respect to the independent variables $\mathbf{x}\in 
X\subseteq\mathbb{R}^n$ up to the order $r$, involving a small parameter $\varepsilon$. 

If one looks for classical Lie point symmetries, in general it is not guaranteed that the infinitesimal generators depend 
on the parameter $\varepsilon$. Nevertheless, the occurrence of terms involving $\varepsilon$ has dramatic effects 
since one loses some symmetries admitted by the unperturbed equation
\begin{equation}
\Delta(\mathbf{x},\mathbf{u},\mathbf{u}^{(r)};0)=0,
\end{equation}
as the following examples clearly show.
\begin{example}
The second order ordinary differential equation
\begin{equation}
\frac{d^2u}{dx^2}+u=0
\end{equation}
admits an eight--dimensional Lie algebra of point symmetries spanned by the vector fields:
\[
\begin{aligned}
&\Xi_1=\frac{\partial}{\partial x}, \quad &&\Xi_2=u\frac{\partial}{\partial u},\\
&\Xi_3=\sin(x)\frac{\partial}{\partial u},\quad &&\Xi_4=\cos(x)\frac{\partial}{\partial u},\\
&\Xi_5=\sin(x)\cos(x)\frac{\partial}{\partial x}-\sin^2(x)u\frac{\partial}{\partial u}, \quad
&&\Xi_6=\cos(2x)\frac{\partial}{\partial x}-\sin(2x)\frac{\partial}{\partial u},\\
&\Xi_7=\cos(x)u\frac{\partial}{\partial x}-\sin(x)u^2\frac{\partial}{\partial u},\quad
&&\Xi_8=\sin(x) u\frac{\partial}{\partial x}+\cos(x) u\frac{\partial}{\partial u}.
\end{aligned}
\]
On the contrary, the equation
\begin{equation}
\frac{d^2u}{dx^2}+u+\varepsilon u^3=0
\end{equation}
admits only the exact symmetry generated by $\Xi_1$.
\end{example}

\begin{example}
The Korteweg--deVries equation
\begin{equation}
\frac{\partial u}{\partial t}+u\frac{\partial u}{\partial x}+\frac{\partial^3u}{\partial x^3}=0
\end{equation}
admits a four--dimensional Lie algebra of exact point symmetries spanned by the vector fields:
\begin{equation}
\Xi_1=\frac{\partial}{\partial t}, \quad \Xi_2=\frac{\partial}{\partial x}, \quad
\Xi_3=t\frac{\partial}{\partial x}+\frac{\partial}{\partial u}, \quad 
\Xi_4=3t\frac{\partial}{\partial t}+x\frac{\partial}{\partial x}-2u\frac{\partial}{\partial u}.
\end{equation}
On the contrary, by considering the Korteweg--deVries--Burgers equation
\begin{equation}
\frac{\partial u}{\partial t}+u\frac{\partial u}{\partial x}+\frac{\partial^3u}{\partial x^3}
-\varepsilon\frac{\partial^2u}{\partial x^2}=0,
\end{equation}
we lose the scaling group and have only three symmetries:
\begin{equation}
\Xi_1=\frac{\partial}{\partial t}, \quad \Xi_2=\frac{\partial}{\partial x}, \quad
\Xi_3=t\frac{\partial}{\partial x}+\frac{\partial}{\partial u}.
\end{equation}
\end{example}

\begin{example}
The $2\times 2$ first order quasilinear system
\begin{equation}
\begin{aligned}
&\frac{\partial u_1}{\partial t}+a_{11}(u_1,u_2)\frac{\partial u_1}{\partial x}
+a_{12}(u_1,u_2)\frac{\partial u_2}{\partial x}=\varepsilon b_1(u_1,u_2),\\
&\frac{\partial u_2}{\partial t}+a_{21}(u_1,u_2)\frac{\partial u_1}{\partial x}
+a_{22}(u_1,u_2)\frac{\partial u_2}{\partial x}=\varepsilon b_2(u_1,u_2),
\end{aligned}
\end{equation}
where $a_{ij}$ and $b_i$ ($i,j=1,2$) are arbitrary functions of the indicated arguments, does not admit an
infinite--dimensional Lie algebra of point symmetries unless it is $\varepsilon=0$. Moreover, also the invariance 
with respect to a homogeneous scaling of the independent variables, \emph{i.e.}, with respect to the Lie generator
\begin{equation}
t\frac{\partial}{\partial t}+x\frac{\partial}{\partial x},
\end{equation} 
is lost when $\varepsilon\neq 0$.
\end{example}

In perturbation theory \cite{Nayfeh} a differential equation involving small terms is often studied by looking for solutions 
in the form
\begin{equation}
\label{expansion_u}
\mathbf{u}(\mathbf{x},\varepsilon)=\sum_{k=0}^p\varepsilon^k \mathbf{u}_{(k)}(\mathbf{x})+O(\varepsilon^{p+1}),
\end{equation}
whereupon the differential equation writes as
\begin{equation}
\Delta\approx \sum_{k=0}^p\varepsilon^k\widetilde{\Delta}_{(k)}\left(\mathbf{x},\mathbf{u}_{(0)},\mathbf{u}^{(r)}_{(0)},
\ldots,\mathbf{u}_{(k)},\mathbf{u}^{(r)}_{(k)}\right)=0.
\end{equation}
Now, let us consider a Lie generator
\begin{equation}
\Xi=\sum_{i=1}^n\xi_i(\mathbf{x},\mathbf{u};\varepsilon)\frac{\partial}{\partial x_i}
+\sum_{\alpha=1}^m\eta_\alpha(\mathbf{x},\mathbf{u};\varepsilon)\frac{\partial}{\partial u_\alpha},
\end{equation}
where we assume that the infinitesimals  depend on the small parameter $\varepsilon$.

According to the results of previous Section, by using the expansion~\eqref{expansion_u} of the dependent variables 
only, we have the following expressions for the infinitesimals:
\begin{equation}
\xi_i\approx\sum_{k=0}^p\varepsilon^k \widetilde{\xi}_{(k)i}, \qquad \eta_\alpha\approx\sum_{k=0}
^p\varepsilon^k\widetilde{\eta}_{(k)\alpha},
\end{equation}
with
\begin{equation}
\begin{aligned}
&\widetilde{\xi}_{(0)i}=\xi_{(0)i}=\left.\xi_i(\mathbf{x},\mathbf{u},\varepsilon)\right|_{\varepsilon=0},\qquad
&&\widetilde{\eta}_{(0)\alpha}=\eta_{(0)\alpha}=\left.\eta_\alpha(\mathbf{x},\mathbf{u},\varepsilon)\right|
_{\varepsilon=0,}\\
&\widetilde{\xi}_{(k+1)i}=\frac{1}{k+1}\mathcal{R}[\widetilde{\xi}_{(k)i}],\qquad &&\widetilde{\eta}_{(k+1)\alpha}=\frac{1}
{k+1}\mathcal{R}[\widetilde{\eta}_{(k)\alpha}],
\end{aligned}
\end{equation}
where, since only the dependent variables are expanded, the recursion operator $\mathcal{R}$ becomes: 
\begin{equation}
\label{R_operator_new}
\begin{aligned}
&\mathcal{R}\left[\frac{\partial^{|\tau|}{f}_{(k)}(\mathbf{x},\mathbf{u}_{(0)})}{\partial u_{(0)1}^{\tau_1}\dots\partial 
u_{(0)m}^{\tau_m}}\right]=\frac{\partial^{|\tau|}{f}_{(k+1)}(\mathbf{x},\mathbf{u}_{(0)})}{\partial u_{(0)1}
^{\tau_1}\dots\partial u_{(0)m}^{\tau_m}}\\
&\phantom{\mathcal{R}\left[\frac{\partial^{|\tau|}{f}_{(k)}(\mathbf{x},\mathbf{u}_{(0)})}{\partial u_{(0)1}
^{\tau_1}\dots\partial u_{(0)m}^{\tau_m}}\right]}
+\sum_{i=1}^m\frac{\partial}{\partial u_{(0)i}}\left(\frac{\partial^{|\tau|} {f}_{(k)}(\mathbf{x},\mathbf{u}_{(0)})}{\partial 
u_{(0)1}^{\tau_1}\dots\partial u_{(0)m}^{\tau_m}}\right)u_{(1)i},\\
&\mathcal{R}[u_{(k)j}]=(k+1)u_{(k+1)j},
\end{aligned}
\end{equation}
for $k\ge 0$,  $j=1,\ldots,m$, $|\tau|=\tau_1+\cdots+\tau_m$.
Thence, we have an approximate Lie generator
\begin{equation}
\Xi\approx \sum_{k=0}^p\varepsilon^k\widetilde{\Xi}_{(k)},
\end{equation}
where
\begin{equation}
\widetilde{\Xi}_{(k)}=\sum_{i=1}^n\widetilde{\xi}_{(k)i}(\mathbf{x},\mathbf{u}_{(0)},\ldots,\mathbf{u}_{(k)})
\frac{\partial}{\partial x_i}
+\sum_{\alpha=1}^m\widetilde{\eta}_{(k)\alpha}(\mathbf{x},\mathbf{u}_{(0)},\ldots,\mathbf{u}_{(k)})\frac{\partial}{\partial u_\alpha}.
\end{equation}

Since we have to deal with differential equations, we need to prolong the Lie generator
to account for the transformation of derivatives. This is done as in classical Lie group analysis of differential equations,
\emph{i.e.}, the derivatives are transformed in such a way the contact conditions are preserved. 
Therefore, we have the prolongations
\begin{equation}
\begin{aligned}
&\Xi^{(0)}=\Xi,\\
&\Xi^{(r)}=\Xi^{(r-1)}+\sum_{\alpha=1}^m\sum_{i_1=1}^n\ldots\sum_{i_r=1}^n\eta_{\alpha,i_1\ldots i_r}\frac{\partial}
{\partial \frac{\partial^r u_\alpha}{\partial x_{i_1}\ldots\partial x_{i_r}}},\qquad r>0,
\end{aligned}
\end{equation}
where
\begin{equation}
\eta_{\alpha,i_1\ldots i_r}=\frac{D \eta_{\alpha,i_1\ldots i_{r-1}}}{D x_{i_r}}-\sum_{k=1}^n\frac{D \xi_k}{D x_{i_r}}
\frac{\partial^r u_\alpha}{\partial x_{i_1}\ldots \partial x_{i_{r-1}}\partial x_k},
\end{equation}
along with the Lie derivative defined as
\begin{equation}
\frac{D}{Dx_i}=\frac{\partial}{\partial x_i}+\sum_{\alpha=1}^m\left(\frac{\partial u_{\alpha}}{\partial x_i}\frac{\partial}
{\partial u_{\alpha}}+\sum_{j=1}^n\frac{\partial^2 u_{\alpha}}{\partial x_i\partial x_j}\frac{\partial}{\partial (\partial 
u_{\alpha}/\partial x_j)}+\cdots\right).
\end{equation}

Of course, in the expression of prolongations, we need to take into account the expansions of $\xi_i$, $\eta_\alpha$ 
and $u_\alpha$ , and drop the $O(\varepsilon^{p+1})$ terms.

\begin{example}
Let $p=1$, and consider the approximate Lie generator
\begin{equation}
\begin{aligned}
\Xi &\approx \sum_{i=1}^n\left(\xi_{(0)i}+\varepsilon\left(
 \xi_{(1)i}+\sum_{\beta=1}^m\frac{\partial \xi_{(0)i}}{\partial u_{(0)\beta}}u_{(1)\beta}\right)\right)\frac{\partial}{\partial x_i}
\\
&+\sum_{\alpha=1}^m\left(\eta_{(0)\alpha}+\varepsilon\left(
 \eta_{(1)\alpha}+\sum_{\beta=1}^m\frac{\partial \eta_{(0)\alpha}}{\partial u_{(0)\beta}}u_{(1)\beta}\right)\right)
 \frac{\partial}{\partial u_\alpha},
 \end{aligned}
\end{equation}
where $\xi_{(0)i}$, $\xi_{(1)i}$, $\eta_{(0)\alpha}$ and $\eta_{(1)\alpha}$ depend on $(\mathbf{x},\mathbf{u}_{(0)})$.
The first order prolongation is
\begin{equation}
\Xi^{(1)}\approx\Xi + \sum_{\alpha=1}^m\sum_{i=1}^n \eta_{\alpha,i}\frac{\partial}{\partial \frac{\partial u_\alpha}{\partial 
x_i}},
\end{equation}
where
\begin{equation}
\begin{aligned}
\eta_{\alpha,i} &= \frac{D}{D x_i}\left(\eta_{(0)\alpha}+\varepsilon\left(
 \eta_{(1)\alpha}+\sum_{\beta=1}^m\frac{\partial \eta_{(0)\alpha}}{\partial u_{(0)\beta}}u_{(1)\beta}\right)\right)\\
 &-\sum_{j=1}^n \frac{D}{D x_i}\left(\xi_{(0)j}+\varepsilon\left(
 \xi_{(1)j}+\sum_{\beta=1}^m\frac{\partial \xi_{(0)j}}{\partial u_{(0)\beta}}u_{(1)\beta}\right)\right)
 \left(\frac{\partial u_{(0)\alpha}}{\partial x_j}+\varepsilon \frac{\partial u_{(1)\alpha}}{\partial x_j}\right),
\end{aligned}
\end{equation}
with the  Lie derivative now defined as
\begin{equation}
\frac{D}{Dx_i}=\frac{\partial}{\partial x_i}+\sum_{k=0}^p\sum_{\alpha=1}^m \left(\frac{\partial u_{(k)\alpha}}{\partial x_i}
\frac{\partial}{\partial u_{(k)\alpha}}
+\sum_{j=1}^n\frac{\partial^2 u_{(k)\alpha}}{\partial x_i\partial x_j}\frac{\partial}{\partial (\partial u_{(k)\alpha}/\partial 
x_j)}+\cdots\right).
\end{equation}
Things go similarly for higher order prolongations.
\end{example}

The approximate (at the order $p$) invariance condition of a differential equation reads
\begin{equation}
\left.\Xi^{(r)}\Delta\right|_{\Delta\approx 0}\approx 0.
\end{equation}
In the resulting condition we have to insert the expansion of $\mathbf{u}$ in order to obtain the determining 
equations at the various orders in $\varepsilon$.

The Lie generator $\widetilde{\Xi}_{(0)}$ is always a symmetry of the unperturbed equations ($\varepsilon=0$); the  
\emph{correction}
terms $\displaystyle\sum_{k=1}^p\varepsilon^k\widetilde{\Xi}_{(k)}$ give
the deformation of the symmetry due to the terms involving $\varepsilon$. 

\begin{remark}
Not all the symmetries of the unperturbed equations are admitted as the zeroth terms of the approximate 
symmetries; the symmetries
of the unperturbed equations that are the zeroth terms of the approximate symmetries are called \emph{stable 
symmetries} \cite{BGI-1989}.  
\end{remark}

\begin{remark}
If $\Xi$ is the generator of an approximate Lie point symmetry of a differential equation, $\varepsilon\Xi$ is a generator 
of an approximate Lie point symmetry too, but the converse is not true in general.
\end{remark}

By the same arguments as in classical Lie theory of differential equations, it remains proved the following result.

\begin{theorem}
The approximate Lie point symmetries of a differential equation are the elements of an approximate Lie algebra.
\end{theorem}

Let us show some simple examples of first order approximate Lie symmetries admitted by differential equations.

\begin{example}
The second order ordinary differential equation
\begin{equation}
\frac{d^2u}{dx^2}+u+\varepsilon u^3=0
\end{equation}
admits a nine--dimensional approximate Lie algebra of first order approximate point symmetries spanned by the vector 
fields:
\[
\begin{aligned}
&\Xi_1=\frac{\partial}{\partial x}, && \Xi_2=\varepsilon u_0\frac{\partial}{\partial u},\\
&\Xi_3=\varepsilon \sin (x)\frac{\partial}{\partial u}, && \Xi_4=\varepsilon \cos(x)\frac{\partial}{\partial u},\\
&\Xi_5=\varepsilon \left(\cos(2x)\frac{\partial}{\partial x}-\sin(2x)u_0\frac{\partial}{\partial u}\right), 
&&\Xi_6=\varepsilon \left(\sin(2x)\frac{\partial}{\partial x}+\cos(2x)u_0\frac{\partial}{\partial u}\right),\\
& \Xi_7=\varepsilon \left(u_0\sin(x)\frac{\partial}{\partial x}+u_0^2\cos(x)\frac{\partial}{\partial u}\right), 
&& \Xi_8=\varepsilon \left(u_0\cos(x)\frac{\partial}{\partial x}-u_0^2\sin(x)\frac{\partial}{\partial u}\right),\\
&\Xi_9=\varepsilon \frac{\partial}{\partial x}.
\end{aligned}
\]
\end{example}

\begin{example}[KdVB equations]
Consider the Korteweg--deVries equation perturbed with the addition of a small dissipative term,
\begin{equation}
\frac{\partial u}{\partial t}+u\frac{\partial u}{\partial x}+\frac{\partial^3u}{\partial x^3}
-\varepsilon\frac{\partial^2u}{\partial x^2}=0.
\end{equation}
The first order approximate symmetries are spanned by the following vector fields:
\begin{equation}
\label{symmetries_kdvb}
\begin{aligned}
&\Xi_1=\frac{\partial}{\partial t}, \qquad \Xi_2=\frac{\partial}{\partial x},\qquad
\Xi_3=t\frac{\partial}{\partial x}+\frac{\partial}{\partial u},\\ 
&\Xi_4=\varepsilon\frac{\partial}{\partial t}, \qquad \Xi_5=\varepsilon\frac{\partial}{\partial x},
\qquad \Xi_6=\varepsilon\left(t\frac{\partial}{\partial x}+\frac{\partial}{\partial u}\right),\\
&\Xi_{7}=\varepsilon\left(3t\frac{\partial}{\partial t}+x\frac{\partial}{\partial x}-2u_0\frac{\partial}{\partial u}\right).
\end{aligned}
\end{equation}

Analogously, the Burgers equation perturbed with the addition of a small dispersive term,
\begin{equation}
\frac{\partial u}{\partial t}+u\frac{\partial u}{\partial x}-\frac{\partial^2u}{\partial x^2}
+\varepsilon\frac{\partial^3u}{\partial x^3}=0,
\end{equation}
admits the first order approximate symmetries spanned by the following vector fields:
\begin{equation}
\label{symmetries_bkdv}
\begin{aligned}
&\Xi_1=\frac{\partial}{\partial t}, \qquad \Xi_2=\frac{\partial}{\partial x},\qquad
\Xi_3=t\frac{\partial}{\partial x}+\frac{\partial}{\partial u},\\ 
&\Xi_4=\varepsilon\frac{\partial}{\partial t}, \qquad \Xi_5=\varepsilon\frac{\partial}{\partial x},
\qquad \Xi_6=\varepsilon\left(t\frac{\partial}{\partial x}+\frac{\partial}{\partial u}\right),\\
&\Xi_{7}=\varepsilon\left(2t\frac{\partial}{\partial t}+x\frac{\partial}{\partial x}-u_0\frac{\partial}{\partial u}\right),\qquad 
\Xi_{8}=\varepsilon\left(t^2\frac{\partial}{\partial t}+tx\frac{\partial}{\partial x}+(x-tu_0)\frac{\partial}{\partial u}\right).
\end{aligned}
\end{equation}
\end{example}

\section{Applications}
\label{applications}
The approximate Lie symmetries can be used to lower the order of ordinary differential equations, as well as to 
compute approximately invariant solutions of partial differential equations. For simplicity, in the following examples we 
take $p=1$, \emph{i.e.}, we consider first order approximate symmetries.
For higher values of $p$ what is only needed is a larger amount of computations. Moreover, since hereafter only scalar differential equations are considered, in order to simplify the notation, the expansion for the unknown 
$u$ is written as $u= u_0+\varepsilon u_1+O(\varepsilon^2)$.

\subsection{Order lowering of an ordinary differential equation}
Here we are going to show how a solvable approximate Lie algebra of symmetries can be used to lower the order of an ordinary 
differential equation.

\begin{example}[Perturbed Blasius equation]
Let us consider the equation
\begin{equation}\label{blasius_perturbed}
\frac{d^3 u}{d x^3}+\frac{1}{2}u\frac{d^2u}{dx^2}+\varepsilon u \frac{du}{dx}=0,
\end{equation}
that for $\varepsilon=0$ is the well known Blasius equation. 
For $\varepsilon\neq 0$,  equation \eqref{blasius_perturbed} admits only one exact Lie point symmetry, say
\begin{equation}
\Xi=\frac{\partial}{\partial x}.
\end{equation}
On the contrary, looking for first order approximate symmetries, equation \eqref{blasius_perturbed}
admits a four--dimensional approximate Lie algebra spanned by the vector fields:
\begin{equation}
\begin{aligned}
&\Xi_1=\frac{\partial}{\partial x},\qquad \Xi_2=\varepsilon\frac{\partial}{\partial x},\qquad 
\Xi_3=\varepsilon\left(x\frac{\partial}{\partial x}-u_0\frac{\partial}{\partial u}\right),\\
&\Xi_4=x\frac{\partial}{\partial x}-u_0\frac{\partial}{\partial u}+\varepsilon\left(\frac{x^2}{3}\frac{\partial}{\partial x}+\left(8-
\frac{2}{3}u_0 x-u_1\right)\frac{\partial}{\partial u}\right).
\end{aligned}
\end{equation}
Let us consider the approximate operators $\widetilde\Xi_1=\Xi_1+\frac{2}{3}\Xi_3$ and $\widetilde\Xi_2=\Xi_4$; it is
$\left[\widetilde\Xi_1,\widetilde\Xi_2\right]\approx \widetilde\Xi_1$, that means that $\widetilde\Xi_1$ and 
$\widetilde\Xi_2$  span an approximate (at first order) solvable two--dimensional Lie subalgebra.
Let us introduce the canonical variables for the operator $\widetilde\Xi_1$, say
\begin{equation}
\widetilde\Xi_1(v)=0,\qquad \widetilde\Xi_1(t)=1,
\end{equation}
whereupon
\begin{equation}
v=u_0+\varepsilon\left(\frac{2}{3}u_0 x+u_1\right),\qquad t=x-\varepsilon\frac{x^2}{3},
\end{equation}
and equation \eqref{blasius_perturbed} becomes
\begin{equation}
\frac{d^3 v}{d t^3}+\frac{1}{2}v\frac{d^2v}{dt^2}-4\varepsilon\frac{d^2v}{dt^2}=0.
\end{equation}
The standard substitution $\displaystyle\frac{dv}{dt}=w(v)$ yields, if $w\neq 0$,
\begin{equation}
\label{eq_1step}
w\frac{d^2w}{dv^2}+\left(\frac{dw}{dv}\right)^2+\frac{1}{2}v\frac{dw}{dv}-4\varepsilon \frac{dw}{dv}=0.
\end{equation}
In terms of the new variables, the first order prolonged operator $\widetilde\Xi_2$  assumes the form
\begin{equation}
\widetilde\Xi_2=(v-8\varepsilon)\frac{\partial}{\partial v}+2w\frac{\partial}{\partial w};
\end{equation}
by computing the new independent and dependent variables $r$ and $s$, respectively, such that
\begin{equation}
\widetilde\Xi_2(r)=1,\qquad \widetilde\Xi_2(s)=0,
\end{equation}
we have
\begin{equation}
\label{Xi2_vars}
r=\log v_0+\varepsilon\left(\frac{v_1}{v_0}-\frac{8}{v_0}\right),\qquad s=\frac{w_0}{v^2}+\varepsilon\left(16 \frac{w_0}
{v^3}+\frac{w_1}{v^2}\right).
\end{equation}
By inserting \eqref{Xi2_vars} in \eqref{eq_1step}, we obtain 
\begin{equation}
s\frac{d^2s}{dr^2}+7s\frac{ds}{dr}+\left(\frac{ds}{dr}\right)^2+\frac{1}{2}\frac{ds}{dr}+6 s^2+s=0.
\end{equation}
Finally, by setting $\displaystyle\frac{ds}{dr}=p(s)$,  the first order ordinary differential equation
\begin{equation}
ps\frac{dp}{ds}+p^2+7ps+\frac{1}{2}p+6s^2+s=0
\end{equation}
is recovered. We notice that the small parameter $\varepsilon$ does not appear explicitly,
but it is somehow hidden inside the involved variables.
\end{example}

\begin{remark}
The reduction of the perturbed Blasius equation to a first order ordinary differential equation where $\varepsilon$ does 
not appear explicitly, as well as the form of the admitted approximate Lie symmetries, suggests us the possibility of 
transforming equation \eqref{blasius_perturbed} 
to the standard Blasius equation. Let us introduce the new independent and dependent variables $t$ and $v$, 
respectively, such that
\begin{equation}
\Xi_4(t)\approx t,\qquad \Xi_4(v)\approx -v,
\end{equation}
whence
\begin{equation}
v=u+\varepsilon\left(\frac{2}{3}x u-8\right),\qquad t=x-\varepsilon\frac{x^2}{3},
\end{equation}
and equation \eqref{blasius_perturbed} writes as
\begin{equation}
\frac{d^3 v}{d t^3}+\frac{1}{2}v\frac{d^2v}{dt^2}=0,
\end{equation}
\textit{i.e.}, the classical Blasius equation.
\end{remark}

\subsection{Approximately invariant solutions of partial differential equations} 
In this Subsection we consider some examples of partial differential equations and compute 
some approximately invariant solutions.
\begin{example}
Consider the nonlinear wave equation
\begin{equation}
\label{nonlinearwave}
\frac{\partial^2 u}{\partial t^2}-\frac{\partial}{\partial x}\left(u^2\frac{\partial u}{\partial x}\right)+\varepsilon \frac{\partial u}
{\partial t}=0.
\end{equation}

The first order approximate symmetries are generated by the following vector fields:
\begin{equation}
\begin{aligned}
&\Xi_1=\frac{\partial}{\partial t}, &&\Xi_2=\frac{\partial}{\partial x},\\
&\Xi_3=\left(t+\varepsilon\frac{ t^2}{6}\right)\frac{\partial}{\partial t}-\left(u_0+\varepsilon\left(u_1+\frac{tu_0}3\right)
\right)\frac{\partial}{\partial u}, 
&&\Xi_4=x\frac{\partial}{\partial x}+(u_0+\varepsilon u_1)\frac{\partial}{\partial u},\\
&\Xi_5=\varepsilon\frac{\partial}{\partial t},&&\Xi_6=\varepsilon\frac{\partial}{\partial x},\\
&\Xi_7=\varepsilon\left(t\frac{\partial}{\partial t}-u_0\frac{\partial}{\partial u}\right), 
&&\Xi_8=\varepsilon\left(x\frac{\partial}{\partial x}+u_0\frac{\partial}{\partial u}\right).
\end{aligned}
\end{equation}

The solutions that result approximately invariant with respect to the Lie generator $\Xi_3$ are such that
\begin{equation}
\left(t+\varepsilon\frac{t^2}{6}\right)\frac{\partial u}{\partial t}=-u_0-\varepsilon\left(u_1+ \frac{tu_0}{3}\right),
\end{equation}
whereupon, insertion of $u=u_0+\varepsilon u_1+O(\varepsilon^2)$, and separation of the coefficients of different
powers of $\varepsilon$, provide the system
\[
\left\{
\begin{aligned}
&t\frac{\partial u_{0}}{\partial t}=-u_0,\\
&t\frac{\partial u_{1}}{\partial t}+\frac{t^2}{6}\frac{\partial u_{0}}{\partial t}=-u_1-\frac{tu_0}{3},
\end{aligned}
\right.
\]
whose solution is
\begin{equation}
\label{representation_nlw}
u_0(t,x)=\frac{U_0(x)}{t}, \qquad u_1(t,x)=\frac{U_1(x)}{t}-\frac{U_0(x)}{6}.
\end{equation}
Substitution of \eqref{representation_nlw} into equation \eqref{nonlinearwave} provides the following reduced system of 
ordinary differential equations:
\begin{equation}
\label{reduced_nlw}
\begin{aligned}
&(U_0^2U_0^{\prime})^\prime-2 U_0 = 0,\\
&(U_0^2 U_1)^{\prime\prime}-2 U_1=0,
\end{aligned}
\end{equation}
where $U_0(x)$ and $U_1(x)$ are functions to be determined, and the prime ${}^\prime$ denotes differentiation with 
respect to $x$.
A solution to the system \eqref{reduced_nlw} is
\begin{equation}
U_0(x)=\pm x, \qquad U_1(x)= \frac{k_1}{x^3}+k_2,
\end{equation}
providing the following approximately invariant solution of \eqref{nonlinearwave}:
\begin{equation}
\label{oursolution_nlw}
u(t,x)=\pm \frac{x}{t}+\varepsilon\left(\frac{k_1}{t x^3}+\frac{k_2}{t}\mp\frac{x}{6} \right).
\end{equation}
\end{example}
\begin{remark}
In \cite{IbragimovKovalev}, equation~\eqref{nonlinearwave} has been analyzed by means of the 
Baikov--Gazizov--Ibragimov approach for approximate symmetries. The solution there obtained 
that is approximately invariant with respect to the first order approximate symmetry generated by
\begin{equation}
\label{Ibragimov_generator}
X= \left(t+\varepsilon\frac{ t^2}{6}\right)\frac{\partial}{\partial t}-\left(u+\varepsilon\frac{tu}{3}\right)\frac{\partial}{\partial u}
\end{equation}
was 
\[
u(t,x)=\pm\left(\frac{x}{t}-\varepsilon\frac{x}{6}\right),
\]
that is less general than \eqref{oursolution_nlw}. The approximate Lie generator 
\eqref{Ibragimov_generator} has the invariants
\[
J_1=x, \qquad J_2=tu+\varepsilon \frac{t^2u}{6},
\]
and the approximately invariant solutions are sought by setting $J_2=\varphi(J_1)$, where the function $\varphi$ is 
determined by satisfying equation \eqref{nonlinearwave}. The solution 
\eqref{oursolution_nlw} could be recovered by setting $J_2=\varphi(J_1,\varepsilon)$ and, then,
taking into account the expansion of $u$.  
\end{remark}

\begin{example}
Consider the Korteweg--deVries--Burgers equation
\[
\frac{\partial u}{\partial t}+u\frac{\partial u}{\partial x}- \frac{\partial^2 u}{\partial x^2}+\varepsilon\frac{\partial^3 u}{\partial 
x^3}=0,
\]
and compute the solutions that are approximately invariant with respect to the Lie generator $\Xi=\Xi_1+c\Xi_2+\Xi_7$ 
(see \eqref{symmetries_bkdv}), where $c$ is a constant, \emph{i.e.},
\[
(1+2\varepsilon t)\frac{\partial u}{\partial t}+\left(c+\varepsilon x\right)\frac{\partial u}{\partial x}=-\varepsilon u_0.
\]
Therefore, inserting $u=u_0+\varepsilon u_1+O(\varepsilon^2)$, we get the system
\[
\left\{
\begin{aligned}
&\frac{\partial u_0}{\partial t}+c\frac{\partial u_0}{\partial x}=0,\\
&\frac{\partial u_1}{\partial t}+c\frac{\partial u_1}{\partial x}+2t\frac{\partial u_0}{\partial t}+x\frac{\partial u_0}{\partial x}
=-u_0,
\end{aligned}
\right.
\]
whose integration provides
\[
\left\{
\begin{aligned}
&u_0(t,x)=U_0(\omega), \\
&u_1(t,x)=\left(\frac{3ct^2}{2}-tx\right)U_0^\prime(\omega)-tU_0(\omega)+U_1(\omega),
\end{aligned}
\right.
\]
where $\omega=x-ct$, and $U_0(\omega)$, $U_1(\omega)$ satisfy the following reduced system of ordinary 
differential equations:
\begin{equation}
\label{reduced_bkdv}
\begin{aligned}
&U_0^{\prime\prime}+\left(U_0-c\right)U_0^\prime=0,\\
&\left(U_1^\prime+(c-U_0)(U_1+U_0^\prime)+\omega U_0\right)^\prime=0,
\end{aligned}
\end{equation}
the prime ${}^\prime$ denoting the differentiation with respect to $\omega$;
by solving system~\eqref{reduced_bkdv}, we find the following first order approximately invariant solution
\[
\begin{aligned}
&u(t,x)=c-2k_1 \tanh \left(k_1 \omega+k_2\right)\\
&\phantom{\frac{1}{8}}+\varepsilon\left(k_4\sech^2(k_1\omega+k_2)+((c+4k_1^2\omega)\cosh((2(k_1\omega+k_2))
\phantom{\frac{1}{8}}\right.\\
&\phantom{\frac{1}{8}}-(4(k_2 c-k_1k_3)(k_1\omega+k_2)+2k_1^2c\omega^2
-32 k_1^4\log(\cosh(k_1\omega+k_2))\\
&\left.\phantom{\frac{1}{8}}+2(k_1(1-k_3)+c(k_1\omega+k_2))\sinh(2(k_1\omega+k_2))\right)
\frac{\sech^2(k_1\omega+k_2)}{8k_1^2},
\end{aligned}
\]
where $k_1$, $k_2$, $k_3$ and $k_4$ are constants.
\end{example}

\section{Conclusions}
\label{conclusions}
In this paper, we proposed a new approach to approximate Lie symmetries of differential equations.
Such a method combines the elegance of the approach by Baikov, Gazizov and Ibragimov \cite{BGI-1989}
with the requirements of perturbative analysis of differential equations. 
Remarkably, it allows to use all the techniques of classical Lie group analysis in an approximate context.

The computational cost in order to determine the approximate symmetries of a differential 
equation using the method proposed here is higher than that of the classical Lie group analysis; nevertheless, a working Reduce 
\cite{Reduce} package doing automatically all the needed work is available
\cite{Relie}.

These approximate symmetries can be used to lower the order of ordinary differential equations as well as to compute 
approximately invariant solutions of partial differential equations. Further applications for deriving approximate 
conservation laws, or local transformations (suggested by the approximate symmetries) mapping differential equations 
to approximately equivalent ones are possible. Moreover, either approximate equivalence transformations 
\cite{Ovsiannikov,Lisle,Meleshko,GOS} for classes of differential equations involving small terms or approximate 
conditional symmetries can be defined \cite{BlumanCole,Arrigo_et_al,Cerniha}. Some of these extensions are currently 
under investigation. 
As a final comment, we observe that this approach can be generalized to include multiple 
scales \cite{Baikov_Ibragimov_ND_2000,Kordyukova_ND2006} in the independent variables in order 
to avoid the occurrence of \emph{secular}--like  terms in the solutions. 
These extensions and generalizations will be the object of forthcoming papers.

\section*{Acknowledgments} 
Work supported by G.N.F.M. of I.N.d.A.M. and by local grants of the University of Messina. The authors thank the referees for their useful comments.

\end{document}